\documentclass[preprint,showpacs,preprintnumbers,amsmath,amssymb]{revtex4}

\usepackage{graphicx}
\usepackage{dcolumn}
\usepackage{bm}

\begin{document}

\title{Spontaneous Symmetry Breaking in General Relativity.\\Brane World Concept.}

\author{Boris E. Meierovich}

\email{meierovich@mail.ru}
\affiliation{P.L.Kapitza Institute for Physical Problems  \\
2 Kosygina str., Moscow 119334, Russia}

 \homepage{http://www.kapitza.ras.ru/people/meierovich/}

\date{\today}

\begin{abstract}
   {Gravitational properties of a hedge-hog type topological defect in
two extra dimensions are considered in General Relativity
employing a vector as the order parameter. All previous
considerations were done using the order parameter in the form of
a multiplet in a target space of scalar fields. The difference of
these two approaches is analyzed and demonstrated in detail.
Regular solutions of the Einstein equations are studied
analytically and numerically. It is shown that the existence of a
negative cosmological constant is sufficient for the spontaneous
symmetry breaking of the initially plain bulk. Regular
configurations have a growing gravitational potential and are able
to trap the matter on the brane. If the energy of spontaneous
symmetry breaking is high, the gravitational potential has several
points of minimum. Identical in the uniform bulk spin-less particles, being trapped within
separate minima, acquire different masses and appear to the
observer on brane as different particles with integer spins.}
\end{abstract}

\maketitle

\section{Introduction}

   The theories of brane world and multidimensional gravity are widely
discussed in the literature. A natural physical concept is that a
distinguished surface in the space-time manifold is a topological
defect appeared as a result of a phase transition with spontaneous
symmetry breaking. The macroscopic theory of phase transitions
allows to consider the brane world concept self-consistently, even
without the knowledge of the nature of physical vacuum. The
properties of topological defects (strings,  monopoles, ...) are
generally described with the aid of multiplet of scalar fields
forming a hedgehog configuration in extra dimensions (see
\cite{Bron 1} and references there in). The scalar multiplet plays
the role of the order parameter. The hedgehog configuration forms
a vector proportional to a unit vector in the Euclidean target
space of scalar fields. This model is self-consistent, but it is
not the only way for generalization of a plane monopole to the
curved space-time.

In a flat space-time there is no difference between a vector and a
hedgehog-type multiplet of scalar fields. On the contrary, in
curved space-time scalar multiplets and vectors are transformed
differently. For this reason in general relativity the two
approaches (a multiplet of scalar fields and a vector order
parameter) give different results which are worth to be compared.
It looks more difficult to deal with a vector order parameter,
and, probably, it is the reason why I couldn't find in the
literature any papers considering phase transitions with a
hedgehog-type vector order parameter in general relativity.

\section{General formulae}

\subsection{Lagrangian}

The order parameter enters the Lagrangian via scalar bilinear
combinations of its derivatives and via a scalar potential $V$
allowing the spontaneous symmetry breaking. If $\phi _{I}$ is a
vector order parameter, then $V$\ should$\ $be$\ $a function of
the scalar $\phi ^{K}\phi _{K}=g^{IK}\phi _{I}\phi _{K},$ and a
bilinear combination of the derivatives is a tensor
\begin{equation}
 S_{IKLM}=\phi_{I;K}\phi_{L;M}.
\end{equation}
Index $_{;K}$ is used as usual for covariant derivatives. There
are three ways to simplify $S_{IKLM}$ into scalars, so the most
general form of the scalar $S,$ formed via contractions of
$S_{IKLM},$ is
\begin{equation}
S=A\left(
\phi_{;K}^{K}\right)^{2}+B\phi_{;K}^{L}\phi_{L}^{;K}+C\phi
_{;K}^{M}\phi _{;M}^{K}, \label{Scalar S}
\end{equation}
where $A,B,$ and $C$ are arbitrary constants. Different
topological defects can be classified by these parameters. In
curved space-time the scalar $S$ depends not only on the
derivatives of the order parameter, but also on the derivatives of
the metric tensor. This is the principle difference between a
vector and a multiplet of scalar fields. \qquad \qquad

The general form of the Lagrangian determining gravitational
properties of topological defects with a vector order parameter is
\begin{equation}
 L\left(\phi_{I},g^{IK},\frac{\partial g_{IK}}{\partial
x^{L}}\right)=L_{g}+L_{d},
\end{equation}
where
\begin{equation}
 L_{g}=\frac{R}{2\kappa^{2}},
 \end{equation} 
 \begin{equation}
L_{d}=A\left( \phi _{;K}^{K}\right)
^{2}+B\phi _{;K}^{I}\phi _{I}^{;K}+C\phi _{;K}^{I}\phi
_{;I}^{K}-V\left( \phi ^{K}\phi _{K}\right) .\label{Ldef}
\end{equation}
$L_{g}$ is the Lagrangian of the gravitational field, $R$ is the
scalar curvature of space-time, $\kappa ^{2}$ is the
(multidimensional) gravitational constant, and $L_{d}$ is the
Lagrangian of a topological defect. Covariant derivation
\begin{equation} \phi _{P;M}=\frac{\partial \phi
_{P}}{\partial x^{M}}-\frac{1}{2}g^{LA}\left( \frac{\partial
g_{AM}}{\partial x^{P}}+\frac{\partial g_{AP}}{\partial
x^{M}}-\frac{\partial g_{MP}}{\partial x^{A}}\right) \phi _{L}
\end{equation}
and razing of indexes $\phi ^{K}=g^{IK}\phi _{I}$ contain $g^{IK}$
and $\frac{\partial g_{IK}}{\partial x^{L}},$ and for this reason
it is convenient to express the Lagrangian as a function of $\phi
_{I},$ $g^{IK},$ and $\frac{\partial g_{IK}}{\partial x^{L}}$.

\subsection{Energy-momentum tensor}

Varying the Lagrangian $L_{d}$ $\left( \ref{Ldef}\right) $ with
respect to $\delta g^{IK}$ and having in mind that
 \begin{equation}
\delta g_{IK}=-g_{KM}g_{IN}\delta g^{NM},
\end{equation}
we get the following expression for the energy-momentum tensor
{\footnote{It differs from (94.4) in  \cite{Landau v2} because
the Lagrangian is considered there as a function of $g^{IK}$ and
$\frac{\partial g^{IK}}{\partial x^{L}}.$ Here and below  $\sqrt{-g}$ stands for  $\sqrt{(-1)^{D-1}g}.$ }}:
\begin{equation}
\begin{array}{c}
T_{IK}=\frac{2}{\sqrt{-g}}\left[ \frac{\partial
\sqrt{-g}L_{d}}{\partial g^{IK}}+g_{QK}g_{PI}\frac{\partial
}{\partial x^{L}}\left( \sqrt{-g}\frac{\partial L_{d}}{\partial
\frac{\partial g_{PQ}}{\partial x^{L}}}\right) \right]
 \label{Tik}
\end{array}
\end{equation}
In the case of the vector order parameter the potential $V\left(
\phi ^{K}\phi _{K}\right) =V\left( g^{IK}\phi _{I}\phi _{K}\right)
$ also undergoes the variation with respect to $\delta g^{IK}$.

It is worth to conduct further derivations with account of
specific properties of particular topological defects.

\section{Global string in extra dimensions}

In my previous papers with Bronnikov (see \cite{Bron 1} and
references there in) we considered global monopoles and strings as
topological defects with the order parameter in the form of a
hedge-hock type multiplet of scalar fields in some flat target
space. The aim of this paper is to describe these defects using
vector order parameter and compare the results.

\subsection{Metric}

The direction of the vector specifies one coordinate, and in the
most simple case the system is uniform and isotropic with respect
to all other coordinates. In our recent paper \cite{Bron 1} we
presented the detailed properties of global strings in two extra
dimensions. For this reason I consider below a topological defect
in the space-time with two extra dimensions. The order parameter
is a space-like vector $(g^{IK}\phi_{I}\phi_{K}<0)$ directed normally from the brane hypersurface and
depending on the only one specific coordinate, namely -- the
distance from the brane. The whole $\left( D=d_{0}+2\right)
$-dimensional space-time has the structure M$^{d_{0}}\times $
R$^{1}\times $ $\Phi ^{1}$ and the metric

\begin{equation}
ds^{2}
=e^{2\gamma \left( l\right) }\eta _{\mu
\nu }dx^{\mu }dx^{\nu }-\left( dl^{2}+e^{2\beta \left( l\right)
}d\varphi ^{2}\right) ,  \label{metric}
\end{equation}
where
$\eta_{\mu \nu }=$ diag $\left( 1,-1,...,-1\right)$ is the
$d_{0}$-dimensional Minkovsky brane metric $\left( d_{0}>1\right)
$, and $\varphi $ is the angular cylindrical coordinate in extra
dimensions. $\gamma $ and $\beta $ are functions of the
distinguished extradimensional coordinate $l$ -- the distance from
the center, i.e. from the brane. $e^{\beta \left( l\right)
}=r\left( l\right) $ is the circular radius. Greek indices $\mu
,\nu ,..$ correspond to $d_{0}$-dimensional space-time on the
brane, and $I,K,...$ -- to all $D=d_{0}+2$ coordinates. The metric
tensor $g_{IK}$ is diagonal, and its nonzero components are
denoted as follows:

\begin{equation}
g_{IK} =\left\{ \begin{array}{l} e^{2\gamma },\quad I=K=0, \\
-e^{2\gamma },\quad 0<I=K<d_{0}, \\  -1,\quad I=K=d_{0}, \\
-e^{2\beta },\quad I=K=\varphi.
\end{array} \right.
\end{equation}


The curvature of the metric on brane due to the matter is supposed
to be much smaller than the curvature of the bulk caused by the
brane formation.

\subsection{Regularity conditions}

If the influence of matter on brane is neglected, then there is no
physical reason for singularities, and the selfconsistent
structure of a topological defect should be regular. A necessary
condition of regularity is finiteness of all invariants of the
Riemann tensor of curvature. The nonzero components of the Riemann
tensor are

\begin{equation}
R_{ \ \ CD}^{AB}= \left\{ \begin{array}{l} -\gamma ^{\prime
2}\left( \delta _{C}^{A}\delta _{D}^{B}-\delta _{D}^{A}\delta
_{C}^{B}\right),\quad A,B,C,D<d_{0}, \\ -\beta ^{\prime }\gamma
^{\prime },\quad A=C=\varphi ,\quad B,D<d_{0}, \\ -\left( \gamma
^{\prime \prime }+\gamma ^{\prime 2}\right) \delta _{D}^{B},\quad
A=C=d_{0},\quad B,D<d_{0}, \\ -\left( \beta ^{\prime \prime
}+\beta ^{\prime 2}\right) ,\quad A=C=d_{0},\quad B=D=\varphi .
\label{R^AB_CD}
\end{array} \right.
\end{equation}

Here prime denotes $d/dl.$ One of the invariants of the Riemann
tensor is the Kretchmann scalar $K=R_{ \ \
CD}^{AB}R_{ \ \ AB}^{CD}$ , which is the sum of all
nonzero $R_{ \ \ CD}^{AB}$ squared. I.e. all the nonzero
components of the Riemann tensor, and namely
\begin{equation}
\gamma ^{\prime },\ \ \gamma^{\prime \prime }+\gamma^{\prime 2}, \
\ \beta^{\prime }\gamma^{\prime }, \ \ \beta^{\prime \prime
}+\beta^{\prime 2}
 \label{Regul conditions}
\end{equation}
must be finite. $r=0$ is a singular point of the cylindrical
coordinate system. The absence of curvature singularity in the
center follows from the last condition $\left( \ref{Regul
conditions}\right) .$ Let
\begin{equation} \beta ^{\prime \prime }+\beta ^{\prime
2}=c<\infty \ \ \ at \ \ l=0. \label{Bet''+Bet'^2=c}
\end{equation}
Integrating $\left( \ref{Bet''+Bet'^2=c}\right) $ in the vicinity
of the center we have \begin{equation} \beta ^{\prime
}=\frac{1}{l}+\frac{1}{3}cl+O\left( l^{3}\right) .
\label{Bet'=1/l+..}
\end{equation}
Relation $\left( \ref{Bet'=1/l+..}\right) $ ensures the correct
$\left( =2\pi \right) $ circumference-to-radius ratio, or,
equivalently, $dr^{2}=dl^{2}$ at $l\rightarrow 0.$ Finiteness of
$\beta ^{\prime }\gamma ^{\prime }$ at $l=0$ is fulfilled if
\begin{equation} \gamma ^{\prime }=O\left( l\right) \label{Gamma'=O(l)}\ \
\end{equation}
at $ l\rightarrow 0,$ or
 smaller.

\subsection{Vector order parameter}

Our aim is to consider the order parameter as a vector in extra
dimensions directed normally from the Minkovsky hypersurface. In
the cylindrical coordinate system of extra dimensions the only
nonzero component of the vector order parameter is\
\begin{equation} \phi _{d_{0}}\equiv \phi .  \label{Fi_d0=Fi}
\end{equation}
In the space-time with the metric $\left( \ref{metric}\right) $\
the covariant derivative \begin{equation} \phi _{I;K}=\delta
_{I}^{d_{0}}\delta _{K}^{d_{0}}\phi ^{\prime }-\frac{1}{2}\delta
_{IK}g^{II}g_{II}^{\prime }\phi  \label{Fi_I;K=}
\end{equation}
is a symmetric tensor: $\phi _{I;K}=\phi _{K;I}.$
For this reason $\phi _{;K}^{I}\phi _{I}^{;K}=\phi _{;K}^{I}\phi _{;I}^{K}$ , and the
Lagrangian $\left( \ref{Ldef}\right) $ takes the form
\begin{equation}
\begin{array}{c}L_{d}=A\left( \phi ^{\prime }+\frac{1}{2}\phi
\sum_{K}g^{KK}g_{KK}^{\prime }\right) ^{2} +\widetilde{B}\left(
\phi ^{\prime 2}+\frac{1}{4}\phi ^{2}\sum_{L}\left(
g^{LL}g_{LL}^{\prime }\right) ^{2}\right) -V\left( -\phi
^{2}\right) \end{array} \label{L_d=...}
\end{equation}
and contains only two arbitrary constants $A$ and
$\widetilde{B}=B+C.$ In $\left( \ref{L_d=...}\right) $ we set
$g^{d_{0}d_{0}}=-1$ in accordance with $\left( \ref{metric}\right)
.$ However one should keep in mind that $\left(
\ref{L_d=...}\right) $ cannot be used in $\left( \ref{Tik}\right)
.$ To derive the energy-momentum tensor $\left( \ref{Tik}\right)
$\ one should use the Lagrangian $\left( \ref{Ldef}\right) ,$ and
set $g^{d_{0}d_{0}}=-1,$ $\left( g^{d_{0}d_{0}}\right) ^{\prime
}=0$\ after differentiation. Nevertheless, the field equation can
be derived using $\left( \ref{L_d=...}\right) $ in the general
formula \begin{equation} \frac{1}{\sqrt{-g}}\left( \frac{\partial
\sqrt{-g}L_{d}}{\partial \phi ^{\prime }}\right) ^{\prime
}-\frac{\partial L_{d}}{\partial \phi }=0.
\label{(dL_d/dFi')'-dL_d/dFi=0}
\end{equation}
In the space-time with metric $\left( \ref{metric}\right) $ the
sums in $\left( \ref{L_d=...}\right) $ are
\begin{equation}\begin{array}{c}
S_{n}=\frac{1}{2^{n}}\sum_{K}\left( g^{KK}g_{KK}^{\prime }\right)
^{n}= d_{0}\gamma ^{\prime n}+\beta ^{\prime n},\quad
n=1,2,.. . \label{Sums}
\end{array}\end{equation}
and the determinant of the metric tensor is \begin{equation}
g=\left( -1\right) ^{D-1}e^{2\left( d_{0}\gamma +\beta \right) }.
\label{g=e^...}
\end{equation}

\subsection{Field equation}

We consider below the case $A\neq 0,$ $\widetilde{B}=0.$ The case
$A=0,$ $\widetilde{B}\neq 0$ will be considered elsewhere.
Substituting $\left( \ref {L_d=...}\right) $ with $A=\frac{1}{2},$
$\widetilde{B}=0$ into $\left( \ref
{(dL_d/dFi')'-dL_d/dFi=0}\right) $ we get the following field
equation in the case of vector order parameter

\begin{equation}
\left[ \phi ^{\prime }+\left( d_{0}\gamma ^{\prime }+\beta
^{\prime }\right) \phi \right] ^{\prime }+\frac{\partial
V}{\partial \phi }=0. \label{Field equation}
\end{equation}
In the case of the multiplet of scalar fields we had \cite{Bron
1}:
\begin{equation} \phi ^{\prime \prime }+\phi ^{\prime }\left(
d_{0}\gamma ^{\prime }+\beta ^{\prime }\right) -\phi e^{-2\beta
}+\frac{\partial V}{\partial \phi }=0. \label{Scalar field eq}
\end{equation}
Unlike (\ref{Scalar field eq}), the field equation (\ref{Field
equation})  doesn't depend directly on $\beta$ (and thus on the
circular radius $r=\ln \beta )$, but instead includes second
derivatives of the metric tensor. In the flat space-time $\gamma
^{\prime }=0,$ $\beta ^{\prime }=\frac{1}{l},$ $\beta ^{\prime
\prime }=-\frac{1}{l^{2}},e^{-2\beta }=\frac{1}{l^{2}},$ \ and
both field equations reduce to
\begin{equation} \phi ^{\prime \prime }+\frac{1}{l}\phi ^{\prime
}-\frac{1}{l^{2}}\phi +\frac{\partial V}{\partial \phi }=0.
\label{Flat field equation}
\end{equation}

\subsection{Energy-momentum tensor}

The energy-momentum tensor $\left( \ref{Tik}\right) $ inevitably
contains second derivatives. However, with the aid of the field
equation $\left( \ref {Field equation}\right) ,$ the second
derivatives can be excluded. The final result of a rather wearing
derivation is
\begin{equation}
 \begin{array}{c}T_{I}^{K}=\frac{1}{2}\delta
_{I}^{K}\left[ \phi ^{\prime }+\left( d_{0}\gamma ^{\prime }+\beta
^{\prime }\right) \phi \right] ^{2} +\delta _{I}^{K}V+\left(
\delta _{I}^{d_{0}}\delta _{d_{0}}^{K}-\delta _{I}^{K}\right)
\frac{\partial V}{\partial \phi }\phi
 \end{array}\label{T ik=1/2Delta^K_I...}
\end{equation}
Unlike the scalar multiplet case, the energy-momentum tensor
$\left( \ref{T ik=1/2Delta^K_I...}\right) $ contains not only the
potential $V,$ but also its derivative $\frac{\partial V}{\partial
\phi }$.

Correctness of $\left( \ref{T ik=1/2Delta^K_I...}\right) $ is
checked by the derivation of the covariant divergence $T_{I;K}^{K}$ (actually $T_{d_{0};K}^{K})$. Again, with the
aid of the field equation $\left( \ref{Field equation}\right) $ we
confirm that
$T_{d_{0};K}^{K}=0.$

\subsection{Einstein equations}

The same way as in \cite{Bron 1} we use the Einstein equations
in the form
\begin{equation}
R_{I}^{K}=\kappa
^{2}\widetilde{T}_{I}^{K},
\end{equation}
where $R_{I}^{K}$ is the Ricci tensor,
\begin{equation}
R_{I}^{K}=\left\{  \begin{array}{c} \delta _{I}^{K}\left[ \gamma
^{\prime \prime }+\gamma ^{\prime }\left( d_{0}\gamma ^{\prime
}+\beta ^{\prime }\right) \right] ,\quad I<d_{0} \\ \delta
_{d_{0}}^{K}\left[ d_{0}\left( \gamma ^{\prime \prime }+\gamma
^{\prime 2}\right) +\beta ^{\prime \prime }+\beta ^{\prime
2}\right] ,\quad I=d_{0} \\ \delta _{\varphi }^{K}\left[ \beta
^{\prime \prime }+\beta ^{\prime }\left( d_{0}\gamma ^{\prime
}+\beta ^{\prime }\right) \right] ,\quad I=\varphi
\end{array} \right.
\end{equation}
and
\begin{eqnarray*} \begin{array}{c}
\widetilde{T}_{I}^{K}=T_{I}^{K}-\frac{1}{d_{0}}\delta _{I}^{K}T
 =-\frac{1}{d_{0}}\delta _{I}^{K}\left[ \phi ^{\prime }+\left(
d_{0}\gamma ^{\prime }+\beta ^{\prime }\right) \phi \right] ^{2}
 -\delta _{I}^{K}\frac{2}{d_{0}}V+\delta _{I}^{K}\left( \delta
_{I}^{d_{0}}+\frac{1}{d_{0}}\right) \frac{\partial V}{\partial
\phi }\phi.  \end{array}
\end{eqnarray*}

In the case of the vector order parameter the set of Einstein
equations
\begin{eqnarray}
\begin{array}{c}\gamma ^{\prime \prime }+\gamma ^{\prime }\left( d_{0}\gamma
^{\prime }+\beta ^{\prime }\right)= \kappa ^{2}\left[
-\frac{1}{d_{0}}\left[ \phi ^{\prime }+\left( d_{0}\gamma ^{\prime
}+\beta ^{\prime }\right) \phi \right]
^{2}-\frac{2V}{d_{0}}+\frac{1}{d_{0}}\frac{\partial V}{\partial
\phi }\phi \right] \end{array} \label{Gamma''+...} \\
\begin{array}{c} d_{0}\gamma ^{\prime\prime }+\beta ^{\prime \prime }+d_{0}\gamma ^{\prime 2}+\beta
^{\prime 2}=  \kappa ^{2}\left[ -\frac{1}{d_{0}}\left[ \phi
^{\prime }+\left( d_{0}\gamma ^{\prime }+\beta ^{\prime }\right)
\phi \right] ^{2}-\frac{2V}{d_{0}}+\left( 1+\frac{1}{d_{0}}\right)
\frac{\partial V}{\partial \phi }\phi \right]\end{array}
\label{do(Gamma''+Gamma'^2)+...} \\
\begin{array}{c}\beta ^{\prime \prime }+\beta
^{\prime }\left( d_{0}\gamma ^{\prime }+\beta ^{\prime }\right)=
 \kappa ^{2}\left[ -\frac{1}{d_{0}}\left[ \phi ^{\prime
}+\left( d_{0}\gamma ^{\prime }+\beta ^{\prime }\right) \phi
\right] ^{2}-\frac{2V}{d_{0}}+\frac{1}{d_{0}}\frac{\partial
V}{\partial \phi }\phi \right] \end{array} \label{Beta''+...}
\end{eqnarray}
consists of three first order equations with respect to $\gamma
^{\prime },$ $\beta ^{\prime },$ and $\phi $ . Both $\gamma $ and
$\beta $ do not enter the equations $\left(
\ref{Gamma''+...}-\ref{Beta''+...}\right) $\ directly, only via
the derivatives. In the case of a scalar multiplet order
parameter, see eq.$\left( 14-16\right) $ in \cite{Bron 1},
$\beta $ enters the Einstein equations directly, and the system of
equations is of the fourth order.

The field equation $\left( \ref{Field equation}\right) $ is not
independent. It is a consequence of the Einstein equations $\left(
\ref {Gamma''+...}-\ref{Beta''+...}\right) $ \ due to the Bianchi
identity.

\subsubsection{First integral}

Excluding the second derivatives $\gamma ^{\prime \prime }$ and
$\beta ^{\prime \prime }$ in the set $\left(
\ref{Gamma''+...}-\ref{Beta''+...}\right) ,$ we get the relation
\begin{equation}
\begin{array}{c}\left( d_{0}\gamma ^{\prime }+\beta ^{\prime
}\right) ^{2}-\left( d_{0}\gamma ^{\prime 2}+\beta ^{\prime
2}\right)=  -\kappa ^{2}\left\{ \left[ \phi ^{\prime }+\left(
d_{0}\gamma ^{\prime }+\beta ^{\prime }\right) \phi \right]
^{2}+2V\right\} , \end{array}  \label{First integral}
\end{equation}
which can be considered as a first integral of the system $\left(
\ref {Gamma''+...}-\ref{Beta''+...}\right) .$

\subsubsection{Further simplification}

The equations $\left( \ref{Gamma''+...}\right) $ and $\left(
\ref {Beta''+...}\right) $ have the same right hand sides.
Extracting one from the other we get the equation
\begin{equation} \left( \gamma ^{\prime }-\beta ^{\prime }\right)
^{\prime }+\left( \gamma ^{\prime }-\beta ^{\prime }\right) \left(
d_{0}\gamma ^{\prime }+\beta ^{\prime }\right) =0,
\label{(Gamma'-Beta')'+...}
\end{equation}
which can be used instead of one of the equations $\left(
\ref{Gamma''+...}\right) $ and $\left( \ref{Beta''+...}\right) $.
With the aid of the relations $\left( \ref{First integral}\right)
$ and $\left( \ref {(Gamma'-Beta')'+...}\right) $ the complete set
of equations can be reduced to a more simple form. Introducing new
functions
\begin{equation} U=\gamma ^{\prime }-\beta ^{\prime
},\ \ W=d_{0}\gamma ^{\prime }+\beta ^{\prime }, \ \ Z=\phi ^{\prime }+W\phi , \label{new functions}
\end{equation}
we get the set of four first order equations
 \begin{eqnarray}
U^{\prime } = -U W   \\ W^{\prime } =\kappa
^{2}\frac{d_{0}+1}{d_{0}}\left( \frac{\partial V}{\partial \phi
}\phi -2V-Z^{2}\right) -W^{2}  \label{W'=...} \\ \phi ^{\prime }
&=&Z-W\phi   \\ Z^{\prime } &=&-\frac{\partial V}{\partial
\phi }.
\end{eqnarray}
Functions $\beta ^{\prime },$ $\gamma ^{\prime },$ and their
combination $S_{2}=d_{0}\gamma ^{\prime 2}+\beta ^{\prime 2}$
$\left( \ref{Sums}\right) $\ are expressed via $U$ and $W$ as
follows:

\begin{equation} \gamma ^{\prime
}=\frac{U+W}{d_{0}+1}\qquad \beta ^{\prime
}=\frac{W-d_{0}U}{d_{0}+1}\qquad
S_{2}=\frac{d_{0}U^{2}+W^{2}}{d_{0}+1}.
\end{equation}

In terms of $U,$ $W,$ and $Z$ the first integral $\left(
\ref{First integral}\right) ,$

\begin{equation}
W^{2}-U^{2}=-\kappa ^{2}\frac{d_{0}+1}{d_{0}}\left\{
Z^{2}+2V\right\} ,
\end{equation}
allows to simplify $\left( \ref{W'=...}\right) $ even more:

\begin{equation} W^{\prime }=\kappa
^{2}\frac{d_{0}+1}{d_{0}}\frac{\partial V}{\partial \phi }\phi
-U^{2}.
\end{equation}

The set of equations
\begin{equation}
\begin{array}{c}
U^{\prime }=-UW \\
W^{\prime }=\kappa
^{2}\frac{d_{0}+1}{d_{0}}\frac{\partial V}{\partial \phi }\phi
-U^{2} \\
\phi ^{\prime }=Z-W\phi \\
Z^{\prime }=-\frac{\partial
V}{\partial \phi }
\end{array}
\label{Convenient set}
\end{equation}
is most convenient for both analytical and numerical analysis.

\subsection{General analysis of equations}

Equations $\left( \ref{Gamma''+...}-\ref{Beta''+...}\right) $ are
invariant against adding arbitrary constants to $\gamma $ and
$\beta .$ Without loss of generality we can set \begin{equation}
\gamma \left( 0\right) =0.  \label{Gamma(0)=0}
\end{equation}
Requirement of regularity in the center dictates the condition
$\left( \ref {Bet'=1/l+..}\right) ,$ and, if we do not consider
configurations with angle deficit (or surplus), we have
\begin{equation}
 r=e^{\beta }=l, \ \  l\rightarrow 0.
\label{r=l at l to 0}
\end{equation}
Integrating $\left( \ref{(Gamma'-Beta')'+...}\right) $ with
boundary conditions $\left( \ref{Gamma(0)=0},\ref{r=l at l to
0}\right) $ we get \begin{equation} \gamma ^{\prime }-\beta
^{\prime }=-e^{-\left( d_{0}\gamma +\beta \right) }.
\label{G'-B'=-e^(d0G+B)}
\end{equation}
It follows from $\left( \ref{G'-B'=-e^(d0G+B)}\right) $ that
$\beta ^{\prime }>\gamma ^{\prime }$ everywhere.

Recall that topological defects, formed as multiplets of scalar
fields \cite {Bron 1}, are of three types. Integral curves can
terminate with:

A) infinite circular radius $r\left( l\right) $ at $l\rightarrow
\infty ;$

B) finite circular radius $r_{\infty }=r\left( \infty \right)
=const<\infty ; $

C) second center $r=0$ at some finite $l=l_{c}.$

In the vector order parameter case the situation is different.
Equation $\left( \ref{G'-B'=-e^(d0G+B)}\right) $ allows to prove
that a regular configuration cannot terminate neither with a
finite value of circular radius $r_{\infty }$\ at $l\rightarrow
\infty ,$ nor in the second center$.$

Suppose for a moment, that $r_{\infty }=const<\infty .$ Then
$\beta ^{\prime }(\infty )=0,$ and $\left(
\ref{G'-B'=-e^(d0G+B)}\right) $ reduces to $\gamma ^{\prime
}=-\frac{1}{r_{\infty }}e^{-d_{0}\gamma }$\ at $l\rightarrow
\infty .$ After integration we get
\begin{equation}
e^{d_{0}\gamma }=\frac{d_{0}}{r_{\infty }}\left( l_{0}-l\right) ,
\end{equation}
$l_{0}$ is a constant of integration. The l.h.s. is obviously
positive, while the r.h.s. becomes negative and infinitely large
at $l\rightarrow \infty .$ Thus $r_{\infty }=const<\infty $ is
impossible.

The second center is also impossible. In the vicinity of the
second center the l.h.s. of $\left( \ref{G'-B'=-e^(d0G+B)}\right)
$ becomes large positive due to $-\beta ^{\prime },\ $and the
r.h.s. remains negative.

We come to the conclusion that regular configurations of
topological defects with the vector order parameter start at the
center $l=0$ and terminate at
 $l\rightarrow \infty $ with infinitely growing circular radius $r\left(
l\right) \rightarrow \infty .$

It follows from the requirement of regularity $\left(
\ref{Gamma'=O(l)}\right) $ that $\gamma ^{\prime }$ $=\gamma
_{0}^{\prime \prime }l$ at $l\rightarrow 0.$ From the first
integral $\left( \ref{First integral}\right) $ we find the
relation between $\gamma _{0}^{\prime \prime },\phi _{0}^{\prime
},$ and $V_{0}:$
\begin{equation}
\gamma _{0}^{\prime \prime }=-\frac{\kappa ^{2}}{d_{0}}\left(
2\phi _{0}^{\prime 2}+V_{0}\right) ,  \label{relation for bound
cond}
\end{equation}
 where $V_{0}$ is the value of the potential at the center $l=0.$ In both
cases (scalar multiplet and vector order parameter) the value
$\phi _{0}^{\prime }=\phi ^{\prime }\left( 0\right) $ is not
restricted by the equations. The difference is that in the scalar
multiplet case $\phi _{0}^{\prime }$ becomes fixed unanimously by
the requirement of regularity, and in the case of vector order
parameter \bigskip $\phi _{0}^{\prime }$ remains a free parameter.

\subsection{Asymptotic behavior}

Condition of regularity requires that $\gamma ^{\prime }$ is
finite everywhere. Within the area of regularity it tends to a
fixed finite value $\gamma _{\infty }^{\prime }$ at $l\rightarrow
\infty $. As soon as $r\left( l\right) \rightarrow \infty $ at
$l\rightarrow \infty ,$ we see from $\left(
\ref{G'-B'=-e^(d0G+B)}\right) $ that $\gamma ^{\prime }-\beta
^{\prime }\rightarrow 0.$ Thus
 $\beta ^{\prime }\left( \infty
\right) =\gamma_{\infty }^{\prime }.$
 The field
$\phi \left( l\right) $ also tends to its finite value $\phi
_{\infty }=\phi \left( \infty \right) .$ Then it follows from the
field equation $\left( \ref{Field equation}\right) $ that
$\frac{\partial V}{\partial \phi }\rightarrow 0$ at $l\rightarrow
\infty ,$ i.e. the regular configuration terminates at an extremum
of the potential $V\left( \phi \right) $. Let $V_{\infty }=V\left(
\phi _{\infty }\right) ,$ $V^{\prime }\left( \phi _{\infty
}\right) =0.$ From the first integral $\left( \ref{First
integral}\right) $ we find the limiting value $\gamma _{\infty
}^{\prime }:$

\begin{equation}
\gamma _{\infty }^{\prime }=\sqrt{-\frac{2\kappa ^{2}V_{\infty
}}{\left( d_{0}+1\right) \left[ d_{0}+\left( d_{0}+1\right) \kappa
^{2}\phi _{\infty }^{2}\right] }}.  \label{Gam'_inf =}
\end{equation}

A necessary condition of existence of regular configurations of
topological defects with the vector order parameter is $V_{\infty
}<0.$

To find the asymptotic behavior of $\phi \left( l\right) $ and
$W\left( l\right) $ we linearize the equations $\left(
\ref{Convenient set}\right) $ at $l\rightarrow \infty :$
\begin{equation}
\phi =\phi _{\infty }+\delta \phi ,\ \ W=\left(
d_{0}+1\right) \gamma _{\infty }^{\prime }+\delta W,
\end{equation}
\begin{equation}
\begin{array}{c}
\delta W^{\prime }=\kappa ^{2}\frac{d_{0}+1}{d_{0}}V_{\infty
}^{\prime \prime }\phi _{\infty }\delta \phi \\ \delta \phi
^{\prime }=\delta Z-\left( d_{0}+1\right) \gamma _{\infty
}^{\prime }\delta \phi -\phi _{\infty }\delta W \\ \delta
Z^{\prime }=-V_{\infty }^{\prime \prime }\delta \phi
\end{array}
\label{Linear set}
\end{equation}
Here primes denote derivatives $d/dl,$ $\left( \delta W^{\prime
}=d \delta W/dl,...\right) ,$ except $V_{\infty }^{\prime \prime
}=\frac{\partial ^{2}V}{\partial \phi ^{2}}\left| _{\phi =\phi
_{\infty }}\right. .$ \ Excluding $\delta Z$ and $\delta W,$ we
get the second order linear homogeneous equation for $\delta \phi:$

\begin{equation}
\delta \phi ^{\prime \prime }+\left( d_{0}+1\right) \gamma
_{\infty }^{\prime }\delta \phi ^{\prime }+\frac{2\kappa
^{2}\left| V_{\infty }\right| V_{\infty }^{\prime \prime
}}{d_{0}\left( d_{0}+1\right) \gamma _{\infty }^{\prime 2}}\delta
\phi =0.
\end{equation}

If the extremum of the potential is minimum $\left( V_{\infty
}^{\prime \prime }>0\right) $\ its nontrivial solution vanishes at
$l\rightarrow \infty :$ \begin{equation} \delta \phi =Ae^{\lambda
_{+}l}+Be^{\lambda _{-}l}, \label{Solutions at l to inf}
\end{equation}
where $A$ and $B$ are constants of integration, and both
eigenvalues \begin{equation} \lambda _{\pm }=-\frac{\left(
d_{0}+1\right) \gamma _{\infty }^{\prime }}{2}\left( 1\mp
\sqrt{1-\frac{8\kappa ^{2}\left| V_{\infty }\right| V_{\infty
}^{\prime \prime }}{d_{0}\left( d_{0}+1\right) ^{3}\gamma _{\infty
}^{\prime 4}}}\right)  \label{eigenvalues}
\end{equation}
are either negative, or have negative real parts. Absence of
growing solutions is the reason why $\phi _{0}^{\prime }$ remains
a free parameter in the vector order parameter case.

The asymptotic behavior of the field $\phi \left( l\right) $ far
from the center is determined by two constant parameters of the
symmetry breaking potential near its extremum, namely $V_{\infty
}$ and $V_{\infty }^{\prime \prime }.$ If the extremum is minimum,
$V_{\infty }^{\prime \prime }>0$, then the expression under the
root can be both positive and negative. So $\phi \left( l\right) $
can tend to $\phi _{\infty }$ either smoothly, or with
oscillations. In the space of physical parameters the boundary
between smooth and oscillating solutions is determined by the
relation \begin{equation} \frac{8\kappa ^{2}\left| V_{\infty
}\right| V_{\infty }^{\prime \prime }}{d_{0}\left( d_{0}+1\right)
^{3}\gamma _{\infty }^{\prime 4}}=1. \label{Boundary between mon
and osc}
\end{equation}

Oscillating behavior of the field $\phi \left( l\right) $ induces
oscillations of $\beta ^{\prime }$ and $\gamma ^{\prime }.$ If
$\gamma ^{\prime }$ changes sign, then $\gamma \left( l\right) $
can have minimums. Remind, that $\gamma $ acts as a gravitational
potential, so the matter can be trapped near the minimums of
$\gamma \left( l\right) $.

Usually $\phi =0$ is a maximum of the potential $V\left( \phi
\right) .$ It is also an extremum, $\partial V/\partial \phi =0$
at $\phi =0$. Regular configurations, starting from the center
$l=0$ with $\phi \left( 0\right) =0, $ can terminate at
$l\rightarrow \infty $ with $\phi _{\infty }=0$ as well. In this
case $V_{\infty }^{\prime \prime }=V^{\prime \prime }\left(
0\right) <0,$ and the linear set $\left( \ref{Linear set}\right) $
reduces to the following asymptotic equation for $\phi \left(
l\right) :$

\begin{equation} \phi ^{\prime \prime }+\left(
d_{0}+1\right) \gamma _{\infty }^{\prime }\phi ^{\prime }-\left|
V_{\infty }^{\prime \prime }\right| \phi =0.
\end{equation}
Its general solution is a linear combination of vanishing and
growing functions:

\begin{equation} \phi =Ae^{-\lambda
_{+}l}+Be^{-\lambda _{-}l},
\qquad \lambda _{\pm }=\frac{\left(
d_{0}+1\right) \gamma _{\infty }^{\prime }}{2}\pm
\sqrt{\frac{\left( d_{0}+1\right) ^{2}\gamma _{\infty }^{\prime
2}}{4}+\left| V_{\infty }^{\prime \prime }\right| },\quad \quad
l\rightarrow \infty .
\end{equation}

Requirement of regularity demands to exclude the growing solutions
from the consideration. It can be done at the expense of  \ $\phi
_{0}^{\prime }.$  Regular solutions terminating at a maximum of
the potential can exist only at some fixed values of \ $\phi
_{0}^{\prime }.$

\subsection{Boundary conditions}

The complete set of equations determining the structure of
topological defect in the case of vector order parameter $\left(
\ref{Gamma''+...},\ref {Beta''+...},\ref{First integral}\right) $
is of the third order with respect to three unknowns $\gamma
^{\prime },\beta ^{\prime },$ and $\phi $. The simple solution is
determined unanimously by the values of these three functions in
any regular point. The center $l=0$ is a singular point of the
cylindrical coordinate system. The condition $\phi \left( 0\right)
=0$ fulfills for both symmetries (high and broken)$.$ $\beta
^{\prime }$ is infinite at $l=0.$ We have to set the boundary
conditions very close to the center, but not exactly at $l=0.$

For numerical analysis it is convenient to deal with a system of
four first order equations solved against the derivatives $\left(
\ref{Convenient set}\right) .$ The symmetry breaking potential
$V\left( \phi \right) $\ enters the equations $\left( \ref
{Convenient set}\right) $ only via its derivative $\frac{\partial
V}{\partial \phi }.$ If we leave only the main terms in the
boungary conditions: $U=-\frac{1}{l},$ $W=\frac{1}{l}$ at
$l\rightarrow 0,$ then we loose any information about the absolute
value of the potential. The value $V_{0}=V\left( 0\right) $
appears in the next approximation. Using the expansion $\left(
\ref{Bet'=1/l+..}\right) $ of $\beta ^{\prime }$ in the vicinity
of the center and the equation $\left(
\ref{G'-B'=-e^(d0G+B)}\right) ,$ we express $c$ via $\gamma
_{0}^{\prime \prime }$ :
\begin{equation} c=-\left(
d_{0}-2\right) \gamma _{0}^{\prime \prime }.
\end{equation}
To preserve the complete information about the symmetry breaking
potential one has to write the boundary conditions at
$l\rightarrow 0$ as follows
\begin{equation}
\begin{array}{c} U=\frac{1}{3}\left(
d_{0}+1\right) \gamma _{0}^{\prime \prime
}l-\frac{1}{l}, \quad   W=\frac{2}{3}\left( d_{0}+1\right)
\gamma _{0}^{\prime \prime }l+\frac{1}{l}, \quad  \phi =\phi
_{0}^{\prime }l, \quad Z=2\phi _{0}^{\prime }.
\end{array} \label{Boundary conditions}
\end{equation}
The values $\gamma _{0}^{\prime \prime },\phi _{0}^{\prime },$ and
$V_{0}$ are not independent. They are connected with each other by
$\left( \ref {relation for bound cond}\right) .$

\subsection{Solutions in case $\frac{\partial V}{\partial \protect\phi }=0$}

If the potential $V=V_{0}$ does not depend on $\phi $, then it
actually plays the role of the cosmological constant $\Lambda
=\kappa ^{2}V_{0}.$ The peculiarity of the vector order parameter
is that the equations $\left( \ref{Convenient set}\right) $ loose
the information about the potential if $\frac{\partial V}{\partial
\phi }\equiv 0.$ $V_{0}$ is present only in the boundary
conditions $\left( \ref{Boundary conditions}\right) .$ The
equations $\left( \ref{Convenient set}\right) $ with
$\frac{\partial V}{\partial \phi }\equiv 0$\ and boundary
conditions $\left( \ref{Boundary conditions}\right) $ have the
following analytic solution \begin{eqnarray*} U
=-\frac{\sqrt{C}}{\sinh \left( \sqrt{C}l\right) }, \quad  W
=\sqrt{C}\coth \left( \sqrt{C}l\right),  \quad  \phi \left( l\right)
=\frac{2\phi _{0}^{\prime }}{\sqrt{C}}\tanh \allowbreak
\frac{\sqrt{C}l}{2},
\end{eqnarray*}
where \begin{equation} C=2\left( d_{0}+1\right) \gamma
_{0}^{\prime \prime }=-\frac{2\left( d_{0}+1\right) }{d_{0}}\left(
2\kappa ^{2}\phi _{0}^{\prime 2}+\Lambda \right) .
\label{C=2d0(do+1)gamma''}
\end{equation}
The solution is regular if $C\geq 0,$ i.e. $\Lambda \leq -2\kappa
^{2}\phi _{0}^{\prime 2}.$ For $g_{00}=e^{2\gamma }$ and
$r=e^{\beta }$ we find

\begin{eqnarray*}
g_{00}\left( l\right)  =e^{2\gamma }=\left( \cosh
\frac{\sqrt{C}l}{2}\right) ^{\frac{4}{d_{0}+1}}, \quad  r\left(
l\right)  =\frac{2\sinh \left( \frac{\sqrt{C}l}{2}\right)
}{\sqrt{C}}\left( \cosh \frac{\sqrt{C}l}{2}\right)
^{-\frac{d_{0}-1}{d_{0}+1}}.
\end{eqnarray*}
The slope $\phi _{0}^{\prime }$ remains arbitrary. If $\phi
_{0}^{\prime }=0$ this solution reduces to the one found earlier
(see \cite{Bron 1} and \cite {Cline})\ for the special case $\phi
\equiv 0.$ The point is that the Einstein equations with a
negative cosmological constant have a nontrivial solution (with a
nonzero order parameter) even without a symmetry breaking
potential.

The necessary condition of regular solutions with broken symmetry
is the existence of extremum points of $V\left( \phi \right) ,$
where $\frac{\partial V}{\partial \phi }=0.$ In case $V=const$ the
condition $\frac{\partial V}{\partial \phi }=0$ is fulfilled
identically, and formally the order parameter $\phi $ can tend to
any  $\phi _{\infty }$ as $\ \ l\rightarrow \infty .$ The
displayed above analytical solution shows that
the
existence of a negative cosmological constant is sufficient for
the symmetry breaking of a uniform plain bulk.

The special case $C=+0$ in $\left( \ref{C=2d0(do+1)gamma''}\right)
$ when $\gamma _{0}^{\prime \prime }=0$ and  \begin{equation} \phi
_{0}^{\prime }=\pm \sqrt{-\frac{\Lambda }{2\kappa ^{2}}}
\label{Fi'0=}
\end{equation}
corresponds to the plain bulk $g_{00}\left( l\right) =1,$ and
$r\left( l\right) =l$.

\subsection{Weak curvature of space-time}

The limit $\kappa ^{2}\rightarrow 0$ is the transition to a flat
space-time. Functions $\beta ^{\prime }$ and $\gamma ^{\prime }$
reduce to $\beta ^{\prime }=l^{-1},$ $\gamma ^{\prime }=0.$ The
field equation $\left( \ref{Field equation}\right) $ reduces to
$\left( \ref{Flat field equation}\right) $, which is the usual
equation for the order parameter in cylindrical coordinates in a
flat space-time. The symmetry breaking potential $V$ is a function
of \ $\phi ^{2},$ so $\frac{\partial V}{\partial \phi }\sim \phi
,$ and $\left( \ref{Flat field equation}\right) $ has a trivial
solution $\phi =0$ corresponding to the symmetric (not broken)
state. The nontrivial solutions, starting with $\phi \left(
0\right) =0,$ $\phi ^{\prime }\left( 0\right) \neq 0$ and
terminating with $\phi =\phi _{m}$ at an extremum of the potential
$\left( \frac{\partial V\left( \phi _{m}\right) }{\partial \phi
}=0\right) ,$ describe the states of broken symmetry. Equation
$\left( \ref{Flat field equation}\right) $ is nonleniar. However,
depending on the form of the potential $V\left( l\right) $\ it can
also have a sequence of nontrivial solutions $\phi _{n}\left(
l\right) ,$ $n=0,1,2,...,$ with zero boundary conditions $\phi
\left( 0\right) =\phi \left( \infty \right) =0$ on both ends$.$
The discrete sequence of derivatives $\lambda _{n}=\phi
_{n}^{\prime }\left( 0\right) $ forms the eigenvalues for the
eigenfunctions $\phi _{n}\left( l\right) .$ Functions $\phi
_{n}\left( l\right) $ change sign $n$ times. The nontrivial
solutions of the field equation with $\phi ^{\prime }\left(
0\right) $ within the interval $\left( \lambda _{n},\lambda
_{n+1}\right) $ change sign $n+1$ times.

The principle difference between the equations $\left( \ref{Field
equation}\right) $ and $\left( \ref{Flat field equation}\right) $
is that the coefficient $\left( d_{0}\gamma ^{\prime }+\beta
^{\prime }\right) $ at $\phi ^{\prime }$ in curved space-time
doesn't vanish at $l\rightarrow \infty .$ If $\phi =\phi _{m}$ is
a minimum of $V\left( \phi \right) $ then $V^{\prime \prime
}\left( \phi _{m}\right) >0,$ and the linearized field equation
$\left( \ref{Flat field equation}\right) $ in case of flat
space-time at $l\rightarrow \infty $ reduces to
\begin{eqnarray*}
\phi ^{\prime \prime }+V^{\prime \prime }\left( \phi _{m}\right)
\left( \phi -\phi _{m}\right) =0
\end{eqnarray*}
and describes non-vanishing oscillations. In curved space-time the
oscillations vanish at $l\rightarrow \infty $ in accordance with
$\left( \ref {Solutions at l to inf}\right) .$

Further detailed analysis is done with the aid of numerical
integration.

\section{Numerical analysis}
\subsection{Regular solutions in the space of parameters}

The numerical integration of equations $\left( \ref{Convenient
set}\right) $ is performed for the ``Mexicam hat'' potential taken
in the same form as in \cite{Bron 1}: \begin{equation}
V=\frac{\lambda \eta ^{4}}{4}\left[ \varepsilon +\left(
1-\frac{\phi ^{2}}{\eta ^{2}}\right) ^{2}\right]  \label{Mexican
hat}
\end{equation}
The potential $\left( \ref{Mexican hat}\right) $ has three
extremum points -- a maximum at $\phi =0$, and two minima at
$\phi =\pm \eta .$ At the limiting values of the order parameter
\begin{eqnarray*} V_{\infty }^{\prime } &=&0,\quad V_{\infty
}^{\prime \prime }=2\eta ^{2},\qquad \phi _{\infty }=\pm \eta \\
V_{\infty }^{\prime } &=&0,\quad V_{\infty }^{\prime \prime
}=-\eta ^{2},\qquad \phi _{\infty }=0.
\end{eqnarray*}

The dimensionless parameter $\varepsilon $ moves the ``Mexican
hat'' up and down. It is equivalent to adding a cosmological
constant. The energy of spontaneous symmetry breaking is
characterized by $\eta ^{2/\left( D-2\right) },$ and
\begin{equation}
a=\frac{1}{\sqrt{\lambda }\eta } \label{scale}
\end{equation}
determines, as usual, the
length scale. In most cases $a$ is associated with the core radius
of a topological defect. Without loss of generality we set $a=1$
in computations. The strength of gravitational field is
characterized by the dimensionless parameter
\begin{equation}
\Gamma =\kappa ^{2}\eta ^{2}.  \label{Parameter Gamma}
\end{equation}

In the case of vector order parameter the state of broken symmetry
is controled by four parameters $d_{0},\varepsilon ,$ $\Gamma ,$
and $\phi _{0}^{\prime }.$ The main difference is that in the
scalar multiplet case regular configurations with given
$d_{0},\varepsilon ,$ and $\Gamma $ existed only for a fixed value
of $\phi _{0}^{\prime }.$ Now the regular configurations with
given $d_{0},\varepsilon ,$ and $\Gamma $ exist within some
interval $0<\phi _{0}^{\prime }\leq \phi _{0 max}^{\prime
}$ with the upper boundary $\phi _{0 max}^{\prime }$
depending on $d_{0},\varepsilon ,$ and $\Gamma .$ This additional
parametric freedom allows to forget about the so called ``fine
tuning'' of the physical parameters$.$

For visual demonstration it is worth to fix $d_{0}=4$\ and one of
the three other parameters. Then the area of existence of regular
solutions can be presented as a map in the plane of two remaining
parameters.


Fig 1. shows the area of regular configurations in the
plane $\left( \varepsilon ,\phi _{0}^{\prime }\right) $ for
$d_{0}=4$ and $\Gamma =1.$ Depending on the values of $\varepsilon
$ and $\phi _{0}^{\prime }$ the order parameter $\phi \left(
l\right) $\ tends to $+\eta ,$ $0,$ or $-\eta $ as $l\rightarrow
\infty .$ The sequence of curves $f_{n}\left( \varepsilon \right)
$ in Fig. 1 are those where $\phi \left( l\right) \rightarrow 0$
as $l\rightarrow \infty .$ They separate the areas with different
signs of $\phi _{\infty }.$ Below the first curve $f_{1}\left(
\varepsilon \right) $ from the bottom, where $0<\phi _{0}^{\prime
}<f_{1}\left( \varepsilon \right) ,$ the order parameter $\phi
\left( l\right) $ doesn't change sign. Between $f_{1}\left(
\varepsilon \right) <\phi _{0}^{\prime }<f_{2}\left( \varepsilon
\right) $ it changes the sign once. In the area $f_{2}\left(
\varepsilon \right) <\phi _{0}^{\prime }<f_{3}\left( \varepsilon
\right) $ it changes the sign twice, and so on. The curves
$f_{n}\left( \varepsilon \right) $ quickly condense to the upper
red curve $f_{\infty }\left( \varepsilon \right) $ as
$n\rightarrow \infty .$ $f_{\infty }\left( \varepsilon \right) $
is the upper boundary of existence of regular solutions (in the
particular case $d_{0}=4$ and $\Gamma =1).$
\begin{figure}  \centering
   \hspace{-2cm}
    \includegraphics{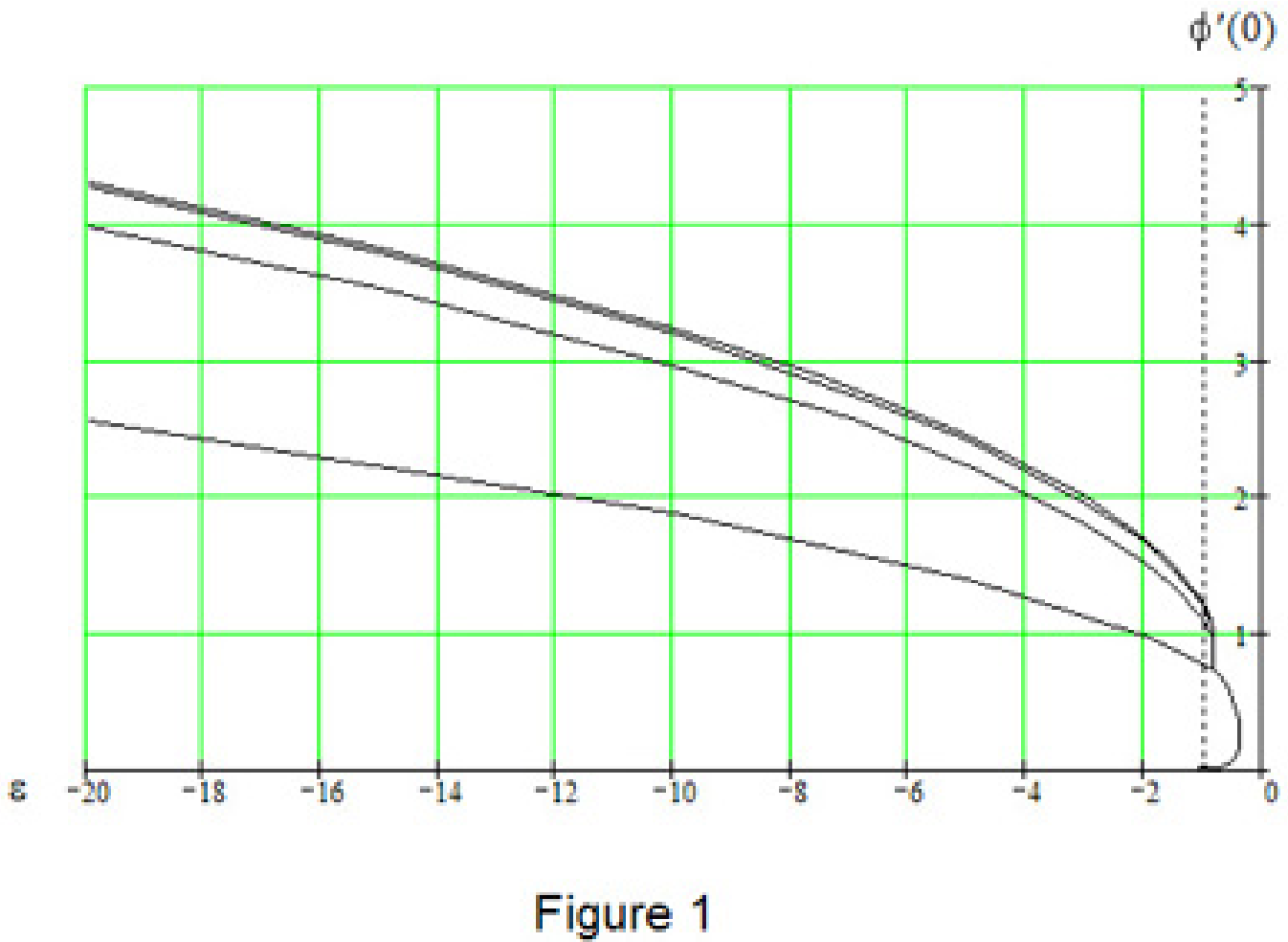}
    \caption{\label{fig:Figure1} The area of regular configurations in the plane $\left( \varepsilon ,\phi
_{0}^{\prime }\right) $ for $d_{0}=4$ and $\Gamma =1.$ The upper curve is the boundary of existence of regular solutions. Other curves separate the regions with different signs of $\phi_\infty. $ They quickly condense to the upper curve. Below the lower curve $\phi(l)$ does not change sign.}
\end{figure}

The curves in Fig. 1 are those where
\begin{equation}
\phi _{\infty }\left( \phi _{0}^{\prime },\varepsilon
,d_{0}=4,\Gamma =1\right) =0.
\end{equation}

 Similar curves can
be shown for fixed $\phi _{0}^{\prime }$ in the plane $\left(
\varepsilon ,\Gamma \right) .$ For instance, the dash line  in Fig.
2 is the first one of the curves $\phi _{\infty }\left( \phi
_{0}^{\prime }=\pm \sqrt{-\frac{\varepsilon +1}{8}},\varepsilon
,d_{0}=4,\Gamma \right) =0,$ where the order parameter tends to
zero at $l\rightarrow \infty $. The value $\phi _{0}^{\prime }=\pm
\sqrt{-\frac{\varepsilon +1}{8}}$ $\left( \ref{Fi'0=}\right) $
corresponds to $\gamma _{0}^{\prime \prime }=0$ in $\left( \ref
{relation for bound cond}\right) .$ It is the case $C=0$ in (\ref
{C=2d0(do+1)gamma''}), so that the symmetry breaking of the plain
bulk is caused completely by the potential $V\left( \phi \right)
,$ and not by the cosmological constant. To the right of the dash
line $\phi \left( l\right) $ does not change the sign.

\begin{figure} \centering
    \includegraphics{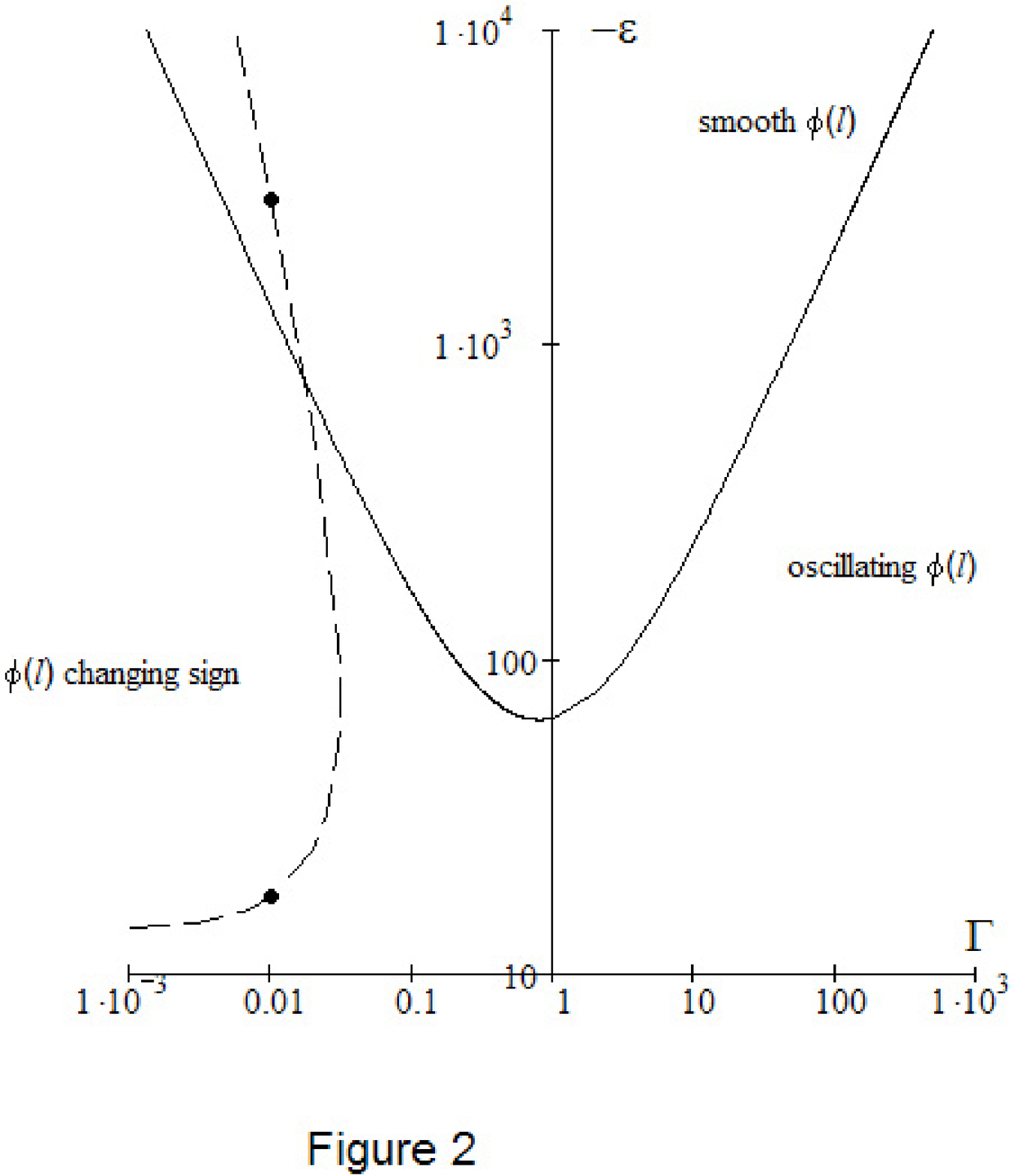}
    \caption{\label{fig:Figure2} Map of regular solutions in the plane $(\Gamma,-\varepsilon)$ for $ \phi
_{0}^{\prime }=\pm \sqrt{-\frac{\varepsilon +1}{8}},
d_{0}=4.$ The red curve separates the regions of smooth (above) and oscillating (below) behavior of the order parameter at $l\rightarrow \infty.$ To the left of the blue curve the order parameter changes sign.}
\end{figure}

For the potential $\left( \ref{Mexican hat}\right) $ the boundary
line $\left( \ref{Boundary between mon and osc}\right) $ between
oscillating and smooth $\phi \left( l\right) $ is \begin{equation}
-\varepsilon _{b}=16\frac{\left( 1+G\right) ^{2}}{G},\quad
G=\frac{d_{0}+1}{d_{0}\Gamma }.  \label{G=}
\end{equation}

%

It is presented in Fig. 2 (solid line). Below the solid line the order
parameter $\phi \left( l\right) $ tends to its limiting value
$\phi _{\infty }$ with damping oscillations (see Fig. 3), and
above this curve -- without oscillations, see Fig. 4. The curves
in Fig. 3 correspond to the close vicinity of the lower black point
on the dash curve in Fig. 2, the curves in Fig. 4 -- to the
vicinity of the upper black point.

\begin{figure}  \centering
    \includegraphics{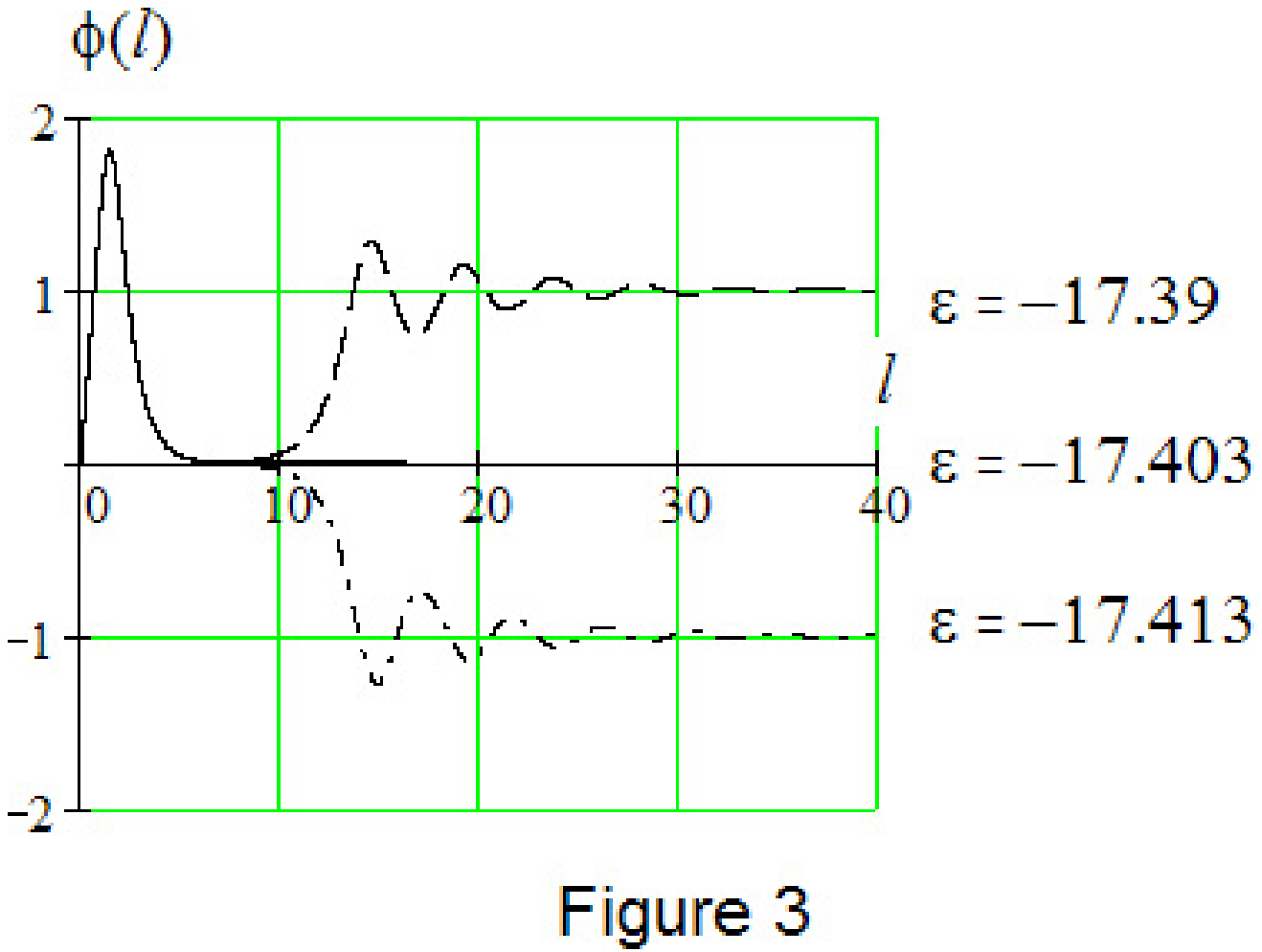}
    \caption{\label{fig:Figure3} Oscillating solutions in the close vicinity of the lower red point on the blue curve in Fig. 2.}
\end{figure}

\begin{figure}\centering
    \includegraphics{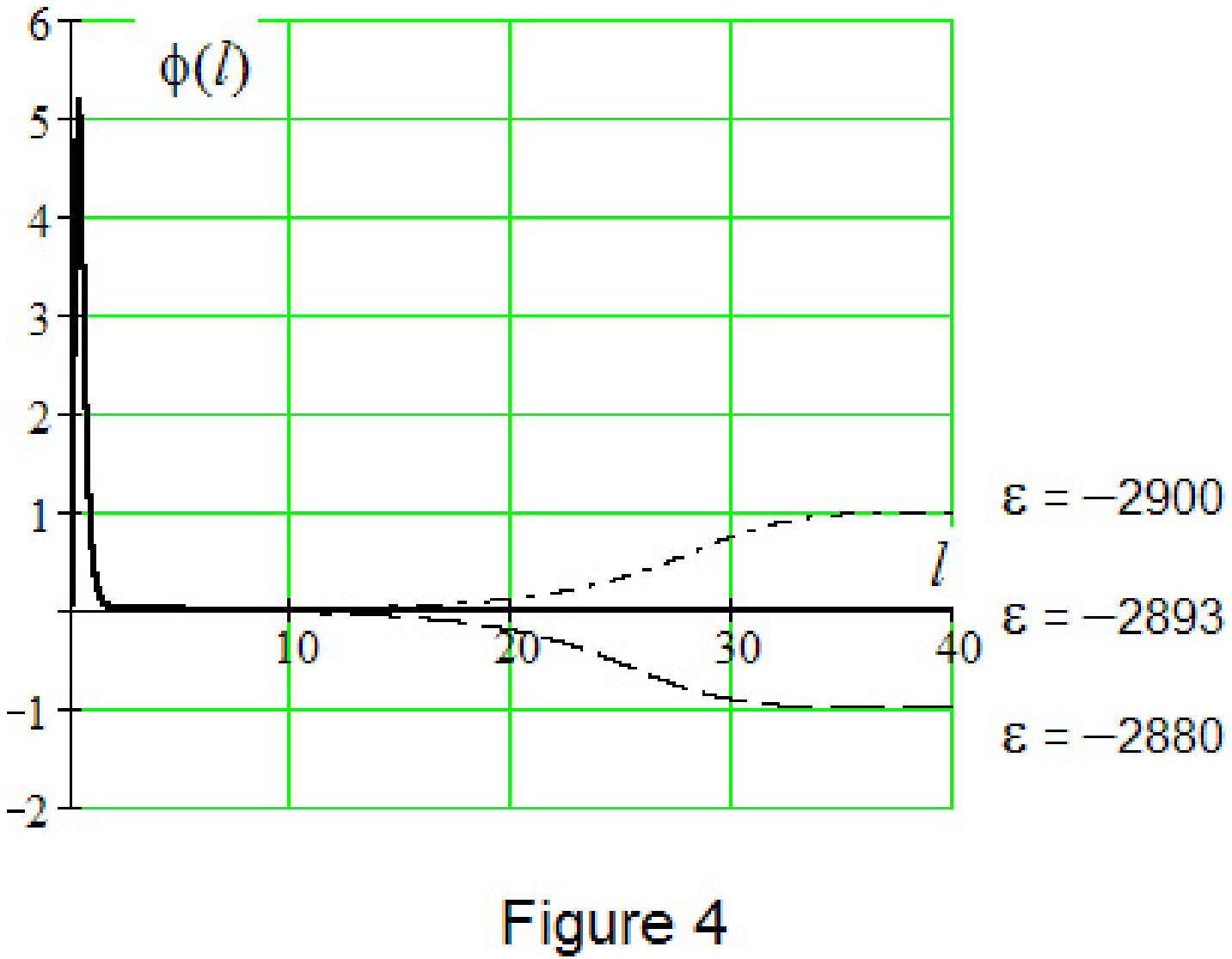}
    \caption{\label{fig:Figure4} Smooth solutions in the close vicinity of the upper red point on the blue curve in Fig. 2.}
\end{figure}

\subsection{Neutral quantum particle in the space-time with metric (\ref{metric}) }

A neutral spinless quantum particle is described by a scalar wave function $\chi $ with the Lagrangian  \\ \begin{equation}
L_{\chi }=\frac{1}{2}g^{AB}\chi _{,B}^{\ast }\chi _{,A}-\frac{1}{2}m_{0}^{2}\chi ^{\ast }\chi .
\end{equation}
In the uniform bulk (while the symmetry is not broken) it is a free particle in the $D$-dimensional space-time with mass $m_{0}$ and spin zero. In the broken symmetry space-time with metric (\ref{metric}) it satisfies the Klein-Gordon equation
\begin{equation}
\frac{1}{\sqrt{-g}}\left( \sqrt{-g}g^{AB}\chi _{,A}\right) _{,B}+m_{0}^{2}\chi =0.
\end{equation}
All coordinates except $x^{d_{0}}=l$ are cyclic variables, and the conjugate momenta are quantum numbers. The wave function in a quantum state is
\begin{equation}
\chi \left( x^{A}\right) =X\left( l\right) \exp \left( -ip_{\mu }x^{\mu }+in\varphi \right) ,
 \end{equation}
 where $p_{\mu }=\left( E,\mathbf{p}\right) $ is the $d_{0}$-momentum within the brane, and $n$ is the integer angular momentum conjugate to the circular extradimensional coordinate $\varphi .$ $X\left( l\right) $ satisfies the equation\cite{Bron 1}
\begin{equation}	
X^{\prime \prime }+WX^{\prime }+\left( p^{2}e^{-2\gamma }-n^{2}e^{-2\beta }-m_{0}^{2}\right) X=0. \label{X''}
\end{equation}
The eigenvalues of  $p^{2}=E^{2}-\mathbf{p}^{2}$ compose the spectrum of
squared masses, as observed in the brane. Quantum number $n$ is the
integer proper angular momentum of the particle.
From the point of view of the observer in the brane it is the internal
 momentum, identical to the spin of the particle.

The equation $\left( \ref{X''}\right) $ takes the form of the Schrodinger equation
\begin{equation}	
y_{xx}+\left[ p^{2}-V_{g}\left( x\right) \right] y=0
\end{equation}
after the substitution

  $dl=e^{\gamma }dx,\qquad X\left( l\right) =y\left( x\right) /\sqrt{f\left( x\right) },$  \qquad
  $f\left( x\right) =\exp \left\{ -\frac{1}{2}\left[ \left( d_{0}-1\right) \gamma +\beta \right] \right\} .$

The gravitational potential
\begin{equation}
V_{g}\left( x\right) =e^{2\gamma }\left( e^{-2\beta }n^{2}+m_{0}^{2}\right) +\frac{1}{2}\frac{1}{\sqrt{f}}\frac{d}{dx}\left( \frac{1}{f^{1/2}}\frac{df}{dx}\right)
 \label{gr.p}
 \end{equation}
 determines the trapping properties of particles to the brane. In terms of $U,W$ and $\phi$  (\ref{new functions})  the dependence of the gravitational potential (\ref{gr.p}) on the distance $l$ is
 \begin{equation}
 V\left( l\right) =e^{2\gamma }\left( e^{-2\beta }n^{2}+m_{0}^{2}\right)
  +\frac{e^{2\gamma }}{4}\frac{\left( d_{0}W-U\right) \left( U+\left( d_{0}+2\right) W\right) }{\left( d_{0}+1\right) ^{2}}+\frac{e^{2\gamma }}{2}\left[ \kappa ^{2}\frac{\partial V}{\partial \phi }\phi +\frac{U\left( W-d_{0}U\right) }{d_{0}+1}\right]. \label{V(l)}
\end{equation}

\subsection{Oscillations}

In terms of $\left( \ref{G=}\right) $\ the eigenvalues $\left(
\ref {eigenvalues}\right) $ are
\begin{equation} \lambda _{\pm
}=-\sqrt{-\frac{\varepsilon }{8\left( G+1\right) }}\left[ 1\pm
\sqrt{1+\frac{16}{\varepsilon G}\left( G+1\right) ^{2}}\right] .
\end{equation}
The
less is $\left| \varepsilon \right| $ the more oscillations
display themselves. In the limiting cases of small and large $\Gamma $ the
frequencies of oscillations
\begin{eqnarray*} \left|
Im \lambda \right| =\left\{ \begin{array}{c} \sqrt{2},\quad
\Gamma \rightarrow 0, \\ \sqrt{2\left( 1+\frac{1}{d_{0}}\right)
\Gamma },\quad \Gamma \rightarrow \infty
\end{array}\right.
\end{eqnarray*}
do not depend on $\varepsilon $ as $l\rightarrow \infty .$

The oscillations of the order parameter $\phi(l)$, see Fig.5, induce the oscillations of the gravitational potential  (\ref{gr.p}).
At $\left| \varepsilon \right| \sim 1$ and
$\Gamma \gg 1$ the gravitational potential
has many points of minimum, see Fig.6.

\begin{figure}\centering
    \includegraphics{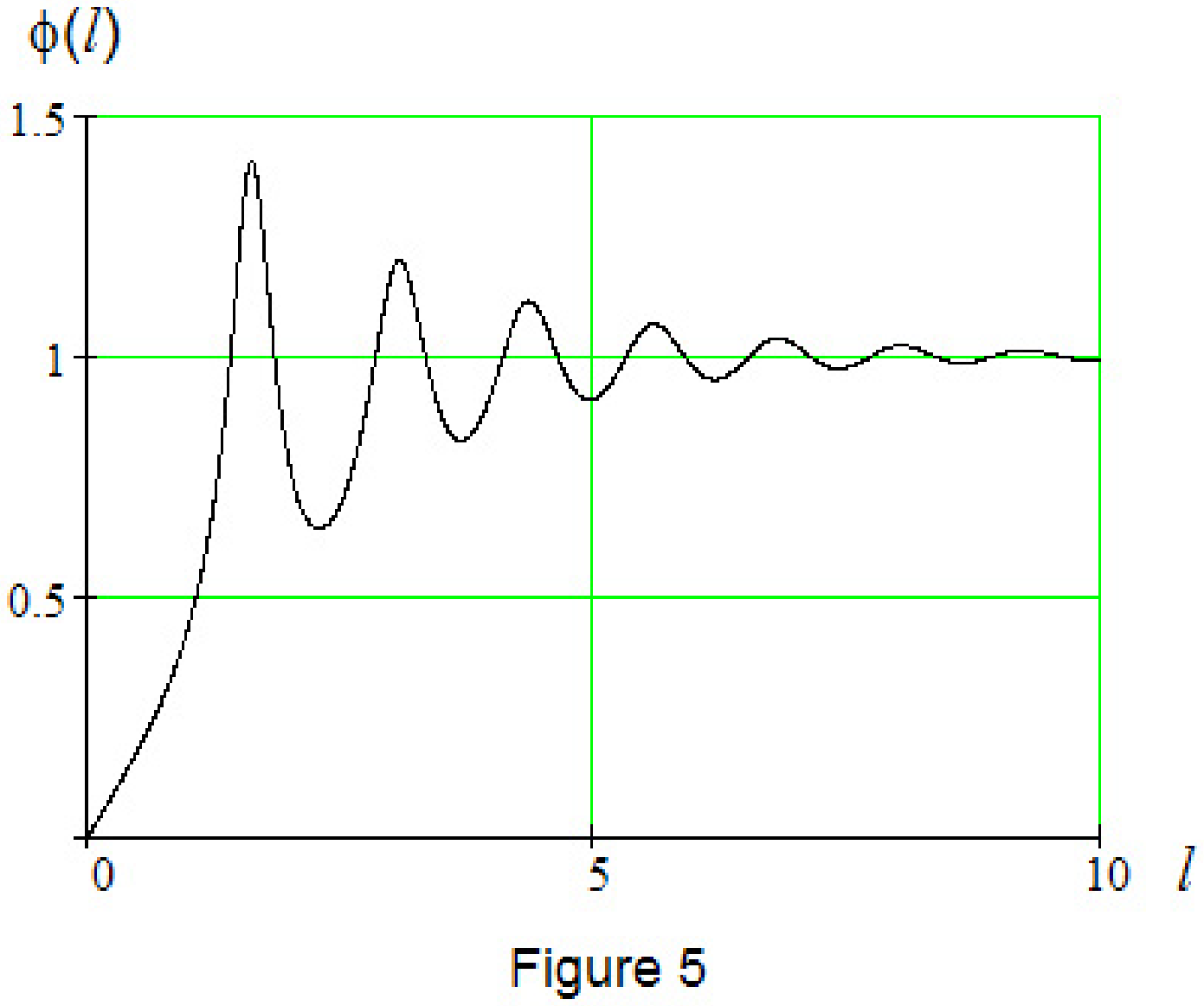}
    \caption{\label{fig:Figure5} A solution with oscillating order parameter $  \phi(l).$ Here $d_{0} =4, \varepsilon=-2, \Gamma=10, \phi_{0}^{\prime }=\sqrt{-\frac{\varepsilon +1}{8}}$.}
\end{figure}
\begin{figure}\centering
    \includegraphics{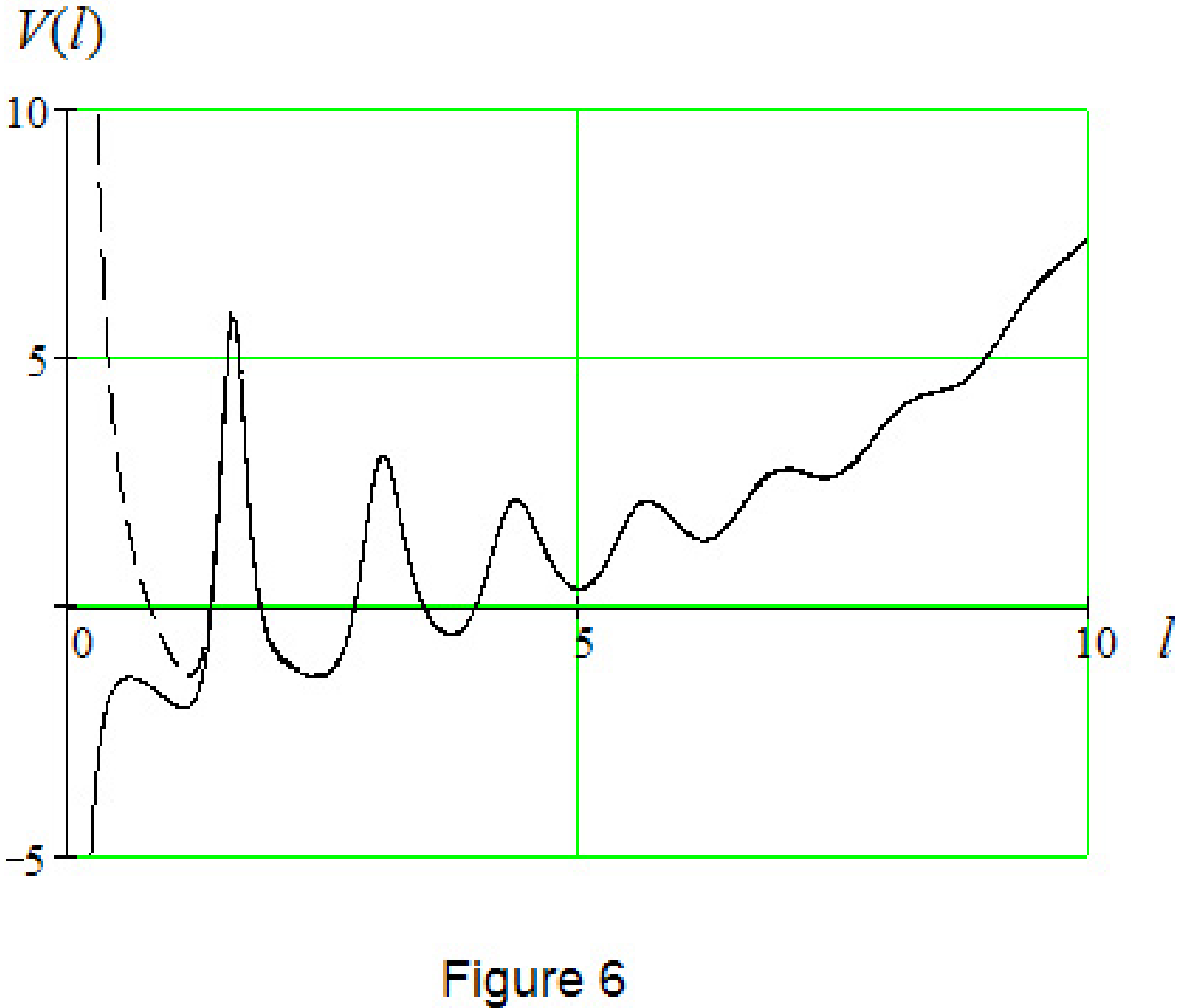}
    \caption{\label{fig:Figure6} Gravitational potential $V_{g}\left( l\right)$ for the same set of the parameters as in Fig.5, $d_{0} =4, \varepsilon=-2, \Gamma=10, \phi_{0}^{\prime }=\sqrt{-\frac{\varepsilon +1}{8}}$. The initial mass of a test particle is set $m_{0}=0$. The red curve corresponds to the angular momentum $n=0$, and the dashed blue one -- to $n=1$. }
\end{figure}

The length scale $a$ (\ref{scale}) remains an arbitrary parameter of the theory. The physical interpretation is different in the limiting cases of large and small $a.$ If $a$ is extremely large, each minimum of the potential $\gamma(l)$ forms its own brane. If the potential barrier is high, the branes are separated from one another.

In the opposite limit, when the scale length $a$ is extremely small, all points of minimum are located within one common brane, and in the spirit of Kalutza-Kline the points of minimum are beyond the resolution of modern devices.

Low energy particles can be trapped by the points of minimum of the potential (\ref{gr.p}). Identical in the bulk neutral spin-less particles,
 being trapped in the different minimum points, acquire different masses and angular momenta. If the scale length $a$ is extremely small, then for
the observer within the brane they appear as different
particles with integer spins.

Most elementary particles have half-integer spins. The simple case of spontaneous symmetry breaking, considered above, can not connect the origin of half-integer spins with extra-dimensional angular momenta.

\end{document}